# Superconductivity under pressure in the two-dimensional van der Waals heavy-fermion metal CeSiI

Tong Shi[1,2#], Wenhao Li[3,4#], Qingxin Dong[1,5 #], Pengtao Yang[1,5], Hanming Ma[1,5], Zhaoming Tian[2], Ningning Wang[1,5], Jianping Sun[1,5], Yoshiya Uwatoko[6,7], Yi-feng Yang[1,5]\*, Bosen Wang[1,5]\*, Hechang Lei[3,4]\*, and Jinguang Cheng[1,5]\*

[1]*Beijing National Laboratory for Condensed Matter Physics and Institute of Physics, Chinese Academy of Sciences, Beijing 100190, China*

[2]*Wuhan National High Magnetic Field Center and School of Physics, Huazhong University of Science and Technology, Wuhan 430074, China*

[3]*Department of Physics and Beijing Key Laboratory of Optoelectronic Functional Materials & Micro-nano Devices, Renmin University of China, Beijing 100872, China*

[4]*Key Laboratory of Quantum State Construction and Manipulation (Ministry of Education), Renmin University of China, Beijing 100872, China*

[5]*School of Physical Sciences, University of Chinese Academy of Sciences, Beijing 100049, China*

[6]*Department of Advanced Materials Science, Graduate School of Frontier Sciences, , University of Tokyo, Kashiwa, Chiba 277-8581, Japan*

[7]*Department of Natural Sciences, Faculty of Science and Engineering, Tokyo City University, Setagaya-ku, Tokyo 158-8557, Japan*

[#]These authors contributed equally to this work.

\*Corresponding authors: bswang@iphy.ac.cn (BSW); hlei@ruc.edu.cn (HCL); yifeng@iphy.ac.cn (YFY); jgcheng@iphy.ac.cn (JGC)

**CeSiI is a newly discovered exfoliable van der Waals (vdW) heavy-fermion metal featured by a long-range antiferromagnetic (AF) order ($T_N \approx 7.5$ K) inside the Kondo coherent state below $T^* \approx 50$ K [1, 2]. To gain a more profound understanding of the intriguing physics of this material and to uncover novel phenomena driven by quantum criticality, it is imperative to construct the phase diagram of CeSiI detailing the evolutions of $T^*$ and $T_N$ as a function of external tuning parameters such as pressure ($P$). In this study, we employ high pressure as an effective tuning knob to investigate this system, thereby generating a comprehensive $T$-$P$ phase diagram of CeSiI. This diagram is characterized by an unusual "V"-shaped nonmonotonic evolution of $T^*(P)$ and the emergence of a superconducting dome with $T_c^{max} \approx 240$ mK upon suppression of AF order at $P_c \approx 6$ GPa, coinciding with the minimum of $T^*(P)$. The close proximity of the superconductivity (SC) to the AF instability and an unusually large upper critical field $B_{c2}(0)$ exceeding 4-7 times the Pauli paramagnetic limit, suggests an unconventional pairing mechanism in CeSiI. Further analyses of normal-state transport properties provide evidence of quantum criticality, i.e., non-Fermi-liquid behavior and divergence of quasiparticle effective mass near $P_c \approx 7$ GPa. Our findings not only establish CeSiI as the first vdW heavy-fermion superconductor but also highlight an unconventional nature for the Kondo coherent state at $T^*$ at ambient pressure, hence opening a new avenue to study the interplay of strong electron correlation, Kondo hybridization, magnetism, and unconventional SC in the vdW heavy-fermion systems [3-6].**

Exfoliable two-dimensional (2D) vdW quantum materials have attracted increasing research interest for exploring dimensionality-driven novel physical phenomena and fabrication of nano-electronic devices [7-15]. The prototype examples include graphene [7, 8], black phosphorus [9,10], transition-metal dichalcogenides [11, 12], and potential monolayer ferromagnets such as $CrI_3$ and $Fe_3GeTe_2$, etc [13-15]. The recent discovery of CeSiI has expanded the vdW quantum materials to heavy-fermion systems, which are characterized by the inheritance of strong electron correlations [2]. As shown in Fig. 1(a, b), CeSiI adopts a layered hexagonal structure with a trigonal $P$-3$m$1 space group. Each block of CeSiI consists of an almost flat honeycomb lattice of Si atoms in the middle, connected to a triangular layer of Ce atoms on both sides, and then covered by iodine atoms at the outermost layers [1, 2]. These CeSiI blocks are stacked alternatively along $c$-axis via weak vdW interactions, facilitating exfoliations of CeSiI flakes down to few-layer limit [2, 5]. Previous studies conducted at ambient pressure (AP) have established bulk CeSiI as a novel vdW heavy-fermion metal with two characteristic transitions upon decreasing temperature, i.e., the formation of Kondo coherence at $T^* \approx 50$ K and an incommensurate AF order at $T_N \approx 7.5$ K. These transitions are manifested by a broad shoulder and a pronounced drop in resistivity $\rho(T)$, respectively [1, 2]. In the recent study, the emergence of coherent hybridization of Ce-4$f$ and conduction electrons at low temperatures was supported by the angle-resolved photoemission spectroscopy (ARPES) and scanning tunnelling microscopy (STM) measurements [2]. However, the full recovery of magnetic doublet entropy $R\ln(2)$ up to 20 K (<< $T^*$) and well retention of Curie-Weiss behavior across $T^*$ are inconsistent with the conventional scenario of Kondo coherent state formation at $T^*$ [2]. This issue is critical as it lays the basis for further exploration of potential topological properties of the material [2, 3, 5]. Moreover, theoretical calculations on CeSiI have highlighted strain-sensitive Kondo hybridization and Ruderman-Kittle-Kasuya-Yosida (RKKY) interaction, positioning it as a prototypical platform for exploring 2D quantum criticality [4]. Preliminary high-pressure (HP) measurements up to 1.6 GPa have shown a downward shift of $T^*$ and $T_N$, but the pressure and temperature ranges are far from an AF quantum critical point (QCP) [2, 4] in the Kondo-magnetic phase diagram [2, 5]. It is thus imperative to extend it further towards the QCP and to explore the relation of the Kondo coherent state and the AF order, as well as possible unconventional SC [2, 5, 16-19]. Here we are motivated to determine the complete evolutions of $T^*$ and $T_N$ as a function of external pressure ($P$) aiming to clarify the unconventional nature of Kondo states and to explore exotic phenomena driven by quantum criticality in CeSiI.

**Pressure evolutions of $T^*$ and $T_N$**

Figure 1(c, d) shows the temperature dependence of resistivity, $\rho(T)$, in a semi-logarithmic scale for a CeSiI single crystal (S1) under various hydrostatic pressures from 2 to 11 GPa down to 50 mK. As seen in Fig. 1(c), $\rho(T)$ at 2 GPa is featured by a broad hump at $T^* \approx 36$ K and a kink-like anomaly at $T_N \approx 5$ K, which are lowered in comparison to those at AP [Extended Data Fig. 1] [1, 2]. It is noted that the shoulder anomaly around $T^*$ at AP is enhanced gradually by pressure and changes into a pronounced hump with increasing pressure. Here the values of $T_N$ and $T^*$ are defined as the points where $d\rho/dT$ peaks out and crosses zero, respectively [Extended Data Fig. 2]. As pressure is applied gradually to 6 GPa, the magnitude of $\rho(T)$ increases monotonically over the entire temperature range, more than doubled at $T^*$, and both anomalies at $T_N$ and $T^*$ are shifted down to lower temperatures continuously, reaching about 3 K and 21 K at 6 GPa, Fig. 1(c). Such a concomitant reduction of $T_N$ and $T^*$ at $P \leq 6$ GPa cannot be understood purely from pressure-enhanced Kondo coherence of the ground state doublet in typical Ce-based heavy-fermion compounds [17, 19, 20, 21], but requires a more intricate mechanism possibly involving crystal field or spin scattering effect [1, 5, 6, 22].

Upon further increasing pressure above 6 GPa, Fig. 1(d), interestingly, the evolution of $\rho(T)$ around $T^*$ is reversed, with its magnitude reduced and $T^*$ shifted to higher temperature progressively, raising up to 110 K at 11 GPa. Meanwhile, the anomaly of $\rho(T)$ at $T_N$ can be hardly defined in the low-temperature region, signaling the complete suppression of AF order at about 6 GPa. These results indicate that the enhancement of Kondo coherence occurs after the low-temperature AF order vanishes.

In addition, we observed a sudden drop of resistivity at pressures above 6 GPa due to the emergence of superconductivity as described below.

The above results are reproduced on another sample CeSiI (S2) measured up to 14 GPa [Extended Data Fig. 3], i.e., its $T^*$ initially decreases to a minimum value of ~20.3 K at 5.5 GPa and then increases monotonically up to ~300 K at 14 GPa, while its $T_N$ is continuously suppressed to ~2.85 K at 5.5 GPa. Besides these similar major features, slight sample-dependent behaviors were noticed for these samples, such as the magnitude of $\rho(T)$ around $T^*$ and the residual resistivity ratio, $RRR \equiv \rho(T^*)/\rho(1.6\ K)$, which are 7.9 (7.4) for S1 (S3) and 1.5 for S2 at AP [Extended Data Fig. 1(b)]. The relatively large $RRR$ value of S1, which is almost two times those (4.3) in Refs. [1, 2], confirms the high-quality of CeSiI crystals used in the present study. We thus focus on this sample in the present study.

**Emergence of superconductivity**

The emergence of superconductivity upon suppression of AF order above about 6 GPa and its evolution with pressure can be clearly seen in Fig. 2(a). At 6.5 GPa, $\rho(T)$ starts to drop from about 0.2 K, signaling the onset of superconducting transition, but zero resistance cannot be achieved down to 0.05 K. By increasing pressure to 7 GPa, the superconducting transition becomes sharper and moves to higher temperatures with $T_c^{onset} \approx 0.24$ K and $T_c^{zero} \approx 0.19$ K, which are defined as the temperatures where $\rho(T)$ starts to deviate from the extrapolated normal-state behavior and reaches zero, respectively, Fig. 2(a). Unexpectedly, the superconducting transition is quickly suppressed upon further applying pressure and becomes undetectable down to 0.05 K at 11 GPa. This indicates that pressure-induced superconductivity in CeSiI exists in a quite narrow pressure range of 6-9 GPa.

The occurrence of superconductivity is reproducible on sample S2 measured in a dilution refrigerator during the decompression process, showing the emergence of superconductivity with $T_c^{onset} \approx 0.14$ K at 5.9 GPa, reaching the maximum of ~ 0.25 K at 6.6 GPa, and then decreasing to ~ 0.15 K at 7.9 GPa [Extended Data Fig. 4(a, b)]. For this sample, zero resistivity was not reached, presumably due to the residual strain upon decompression from 14 GPa and/or the relatively low sample quality (e.g., disorder, non-stoichiometry) with a smaller $RRR$ value shown above.

To gain further insights into the superconducting properties of CeSiI, we studied the effect of magnetic field, $B$, on the superconducting transition by measuring low-$T$ $\rho(T)$ under various magnetic fields at $P$ = 7, 7.5 and 8 GPa, Fig. 2(b) and Extended Data Fig. 5, and $\rho(B)$ at 0.08 K at 7 and 7.5 GPa, Fig. 2(c). As can be seen, the superconducting transition is shifted gradually to lower temperatures with increasing $B$, and the upper critical field, $B_{c2}$, at 0.08 K decreases from ~ 2.4 T at 7 GPa to ~ 1.2 T at 7.5 GPa. To quantify the evolution of zero-temperature upper critical field $B_{c2}(0)$ under pressure, we chose $T_c^{onset}$ to mark the superconducting transition and plot the temperature dependence of $B_{c2}(T)$ in Fig. 2(d), which can be well described by using the Werthamer-Helfand-Hohenberg (WHH) model [23]. The best fits, shown by the dashed lines, yield $B_{c2}(0) \approx 3.03$ T, 1.66 T and 0.98 T for 7, 7.5 and 8 GPa, respectively. These values are about 4-7 times the Pauli paramagnetic limit $B_P = 1.84T_c$, implying an unconventional nature for the observed superconductivity as discussed below.

*T-P* **Phase diagram**

Based on the above results, we construct the *T-P* phase diagram of CeSiI, Fig. 3, which reveals an intriguing concomitant evolution of $T^*(P)$ and $T_N(P)$ as well as the dome-shaped $T_c(P)$. With increasing pressure, both $T^*(P)$ and $T_N(P)$ are initially reduced concurrently: $T^*(P)$ reaches a minimum of 20 K at about 6 GPa and then increases quickly with pressure, while $T_N(P)$ vanishes at the same pressure, signaling the occurrence of pressure-driven AF instability [16, 19-21]. Meanwhile, a dome-like superconducting $T_c(P)$ with a maximum of 0.24 K at ~7 GPa emerges in a narrow pressure range of 6-9 GPa. As shown below, the superconducting dome is developed from a non-Fermi-liquid normal state sandwiched by Fermi-liquid phases on both sides in the *T-P* phase diagram.

While the emergence of superconductivity around the AF QCP closely resembles many other heavy-fermion superconductors and implies an unconventional pairing mechanism [16-19], the unusual nonmonotonic evolution of $T^*(P)$ differs from most Ce-based heavy-fermion materials [17, 20, 21]. Rather, the overall phase diagram is reminiscent of pressurized CeRhIn$_5$ [24], where the initial decrease of $T^*(P)$ has been ascribed to some effects associated with AF spin fluctuations [25] while the *f*-electrons remain localized in this regime [26, 27]. Such a scenario seems to be inconsistent with previous STM and ARPES observations of the Kondo coherent state below $T^*$ at AP for CeSiI [2]. A closer inspection of specific heat and magnetic susceptibility measurements suggests that the *f*-electrons are indeed localized deep below $T^*$ in CeSiI at AP, because the magnetic entropy is fully recovered at about 20 K [2] and the magnetic susceptibility grows continuously across $T^*$ following the Curie-Weiss behavior [Extended Data Fig. 1(c)], with an effective moment $\mu_{eff} = 2.47(1)$ $\mu_B$/f.u. close to the effective moment of free $Ce^{3+}$ ions [1, 5]. Moreover, only light carriers with effective masses comparable to that of a bare electron are observed via the SdH measurements [2]. These experimental facts together with the similar $T^*(P)$ between CeSiI and CeRhIn$_5$ strongly suggest that the Kondo coherent state around $T^*$ cannot be understood simply from the Kondo coherence of the Ce-4*f* ground state doublet but requires more studies.

**Evidence of quantum criticality**

To gain further insight into the AF QCP of CeSiI, we performed quantitative analyses on the resistivity just above $T_c$ by using the power-law formula, viz., $\rho(T) = \rho_0 + AT^n$, where $\rho_0$ is the residual resistivity. The exponent $n = 2$ indicates Fermi-liquid behavior with the quadratic temperature coefficient $A$ proportional to the effective mass of carriers, i.e., $A \propto (m^*/m_0)^2$. We first plot the low-temperature $\rho(T)$ data in the form of $\rho$ versus $T^2$, Fig. 4(a) and Extended Data Fig. 6(b), from which the quadratic-temperature coefficient $A$ can be extracted as the initial slope of a linear fitting to each curve while the intercept gives $\rho_0$.

Fig. 4(b, c) displays the pressure dependencies of $\rho_0$ and $A$, both of which exhibit divergent behaviors around $P_c \approx 7$ GPa, where the superconducting dome reaches the maximum. Notably, the quadratic coefficient $A(P)$ is enhanced by more than an order of magnitude from 2 GPa to $P_c$ and then it is reduced quickly to nearly zero at $P \geq 9$ GPa. Such divergent behavior near $P_c$ can be described by a power-law fitting to $A(P) \propto (P-P_c)^{-\beta}$, which yields $\beta^- \approx 0.5$ and $P_c^- \approx 6.71$ GPa for $P < P_c$ and $\beta^+ \approx 1.25$ and $P_c^+ \approx 6.57$ GPa for $P > P_c$, respectively. Such critical behaviors are characteristic of quantum criticality [28, 29], indicative the divergence of effective electron mass. This is further substantiated by analyzing the upper critical field near $P_c$ as $m^*/m_0 \propto [dB_{c2}(T)/T_c dT|_{T=T_c}]^{1/2}$, which shows an enhancement to 8.5 at 7 GPa, 7.5 at 7.5 GPa, and 7.9 at 8 GPa, respectively. The positive correlation between $A(P)$ and $T_c(P)$ suggests that enhanced $m^*$ near the AF QCP due to spin fluctuations should be responsible for the dome-like superconducting diagram. Meanwhile, it is noteworthy that our fits of $A(P) \propto |P-P_c|^{-\alpha}$ yield different exponents for the pressure regions below and above $P_c$. Such an asymmetry of $A(P)$ becomes apparent when the two sides correspond to different magnetic ground states or even Fermi surface structures [26].

As shown in Fig. 4(a), a deviation from the Fermi-liquid behavior can be clearly observed near $P_c$. To extract the exponent $n$, we performed a polynomial fitting to $\rho(T) = \rho_0 + A'T^n$ for $T_c < T < 2$ K and plotted the obtained $n(P)$ in Fig. 4(d). As can be seen, the exponent $n$ initially retains ~2 within the pressure range of 2-6 GPa, decreases to a minimum value of 1 around $P_c \approx 7$ GPa, and then recovers to ~ 2 above 11 GPa. The emergence of non-Fermi-liquid behavior near $P_c$ can be clearly visualized in the color contour-plot of $n(T, P)$ superimposed on the $T$-$P$ phase diagram in Fig. 3, which is another hallmark of quantum criticality with enhanced critical spin fluctuations near the AF QCP [19, 30-34]

Such $T$-linear resistivity has been observed near the QCP in many heavy-fermion systems, but ascribed to different mechanisms, such as the spin-density-wave (SDW) in 2D or other more exotic scenarios [34]. Which scenario applies to CeSiI is closely associated with the relation of the AF order and the Kondo coherent state. If the AF order originates from Fermi surface instability of the coherent heavy-fermion state, the QCP belongs to the SDW type as reported in CeCoIn$_5$ [35]. On the other hand, if the $f$-electrons remain localized and the AF order is destroyed accompanying with sudden jump of the Fermi surface size, the QCP may be of the Kondo breakdown or other unconventional type, as suggested for CeRhIn$_5$ [36] and YbRh$_2$Si$_2$ [37]. The resemblance of the $T$-$P$ phase diagram between CeSiI and CeRhIn$_5$ suggests that they might belong to the same category. However, frustration might extend the QCP into a finite region with an intermediate spin-liquid-type ground state, as reported previously in the distorted kagome Kondo lattice CePdAl [38]. In this regard, the frustration-driven criticality or competing exchange interactions are credible alternative explanations considering the incommensurate magnetic order and the triangular-lattice motif in CeSiI. More elaborate investigations on the Fermi surface reconstruction are necessary to establish this issue, and reveal other exotic properties around the QCP, potentially arising from the interplay of the Kondo effect, magnetic frustration, superconductivity, and two-dimensionality.

**Discussions**

The discovery of superconductivity near the AF QCP of 2D vdW heavy-fermion metal CeSiI is important considering the following aspects. At first, it enriches the family of heavy-fermion superconductors and provides a new platform for studying unconventional superconducting mechanisms [19, 30-34]. Secondly, the reduction of lattice dimensionality usually enhances quantum fluctuations near magnetic QCPs, which may be beneficial for higher $T_c$ in heavy-fermion superconductors. On this basis, the dimensional tunability can provide new approaches and perspectives for the exploration or study of heavy-fermion superconducting materials. Thirdly, it has established a bridge between 3D heavy-fermion and 2D quantum materials, providing an important approach for studying the emergent dimensionality-limited heavy fermion quantum states via mechanical exfoliation, stress, surface doping, etc. [2, 4]. We have repeated the electrical transport measurements for CeSiI (S4) ($RRR \approx 4.3$) under various pressures. We find that the evolution of $T_N$ and $T^*$ are basically consistent with the results for CeSiI (S1, S2). Interestingly, the anisotropic upper critical field $B_{c2}(0)$ determined by the WHH model for the $B//ab$, ~ 1.58 T for CeSiI (S4) at 7.0 GPa, is very close to the $B_{c2}(0)$ ~ 1.66 T of $B//c$ at 7.5 GPa for CeSiI (S1) in Figure 2d. The observation indicates the possible pseudo-3D superconductivity in the 2D van der Waals structure of CeSiI under pressure. Many theoretical studies including topological magnetism and AF Chern insulator in the monolayer CeSiI have already been proposed as the forefront [1, 3, 4].

Recently, Pasupathy *et al.* have reported that CeSiI exhibits nodal Kondo hybridization as a consequence of its low dimensionality and Ce-site local symmetry [39]. We believe that there are three main aspects to its impact on superconductivity and the AF QCP of CeSiI: firstly, such nodal hybridization may favor anisotropic superconducting state with exotic pairing mechanism and constrain conventional *s*-wave pairing symmetries; secondly, it may induce anisotropic RKKY interaction, resulting in strong magnetic frustration and affecting the nature of the AF QCP; thirdly, it may cause the Fermi surface collapse in a more complex manner (strongly angle dependent) near the QCP. These will possibly lead to new phenomena that have not been explored before and make the AF QCP and superconductivity more intriguing. Hence, our discovery will further stimulate experimental explorations and theoretical calculations, e.g., neutron diffraction to analyze magnetic structures and theoretical calculations to reveal electronic band structures and quantum critical properties.

As noted above, the observation of large $B_{c2}(0)$ that exceeds 4-7 times the conventional Pauli paramagnetic limit $B_P$ near $P_c$, Fig. 4(c), implies an unconventional nature for the observed superconductivity, as found in some Ce-based heavy-fermion materials [40-44]. One possible origin for such a large $B_{c2}(0)$ is the reduction of the $g$ factor, which is related to the interaction-induced mass enhancement [45], as found in $CeSb_2$ [44] and $CeCoIn_5$ [46]. As discussed above, CeSiI shows a remarkable mass re-normalization near $P_c$, which may lead to the high $B_{c2}(0)$. According to Ref. [44], we can estimate the electronic Sommerfeld coefficient $\gamma = 0.85$ J mol$^{-1}$ K$^{-2}$ from the upper critical field at 7 GPa, viz., $\gamma = C(dB_{c2}(T)/T_c dT|_{T=T_c})^{1/2}$, where $C$ is prefactor 0.1 J mol$^{-1}$ K$^{-1}$ T$^{-1/2}$, and $dB_{c2}(T)/T_c dT|_{T=T_c} \approx 71.6$ T K$^{-1}$ at 7 GPa. The obtained $\gamma$ value at 7 GPa exhibits about 7 times enhancement compared to that of 0.125 J mol$^{-1}$ K$^{-2}$ at AP. At the same time, we can also estimate the $\gamma$ value at 7 GPa from the Kadowaki-Woods ratio, i.e., $\gamma_{KW} \approx (A/R_{KW})^{0.5} \approx 0.74$ J mol$^{-1}$ K$^{-2}$ based on the $T^2$-coefficient $A(7 \text{ GPa}) \approx 5.529$ μΩ cm K$^{-2}$ and the universal value $R_{KW} \approx 10$ μΩ cm mol$^2$ K$^2$ J$^{-2}$ for many heavy-fermion compounds. As can be seen, the estimated $\gamma$ values are consistent with each other, indicating its reliability. Such a huge $\gamma$ value further confirms the heavy-fermion features and mass enhancement of CeSiI near the AF QCP.

Another possible reason for high $B_{c2}(0)$ is due to the strong-coupling effect, which could enhance superconducting energy gap Δ. This scenario has successfully explained the large $B_{c2}(0)$ of $UBe_{13}$ [47]. The third possibility leading to high $B_{c2}(0)$ is the mixture or transition between even-singlet and odd-triplet pairing states, as proposed in noncentrosymmetric heavy-fermion superconductors such as $CeCoGe_3$, $CeRhSi_3$, and $CeIrSi_3$, as well as $CeRh_2As_2$ which has a global inversion symmetry but locally noncentrosymmetric environment at Ce sites [42, 43, 48]. Similar to $CeRh_2As_2$, CeSiI is centrosymmetric but Ce atoms are located at locally noncentrosymmetric sites. Whether the strong coupling superconductivity or locally noncentrosymmetric effect exists in CeSiI is worth more detailed theoretical and experimental investigations in the future. At the same time, the orbital limiting field $B_{c2}^{orbit}(0)$ is commonly derived from the slope of the determined $B$-$T$ phase boundary at $T_c$, which is given as $B_{c2}^{orbit}(0) = -0.73 dB_{c2}(T)/T_c dT|_{T=T_c}$ in the clean limit. The relative importance of the orbital and spin-paramagnetic effects can be described by the Maki parameter $\alpha = \sqrt{2} B_{c2}^{orbit}(0)/B_P$. In this work, the estimated $\alpha$ are 9.63 at 7 GPa, 6.09 at 7.5 GPa and 5.13 at 8 GPa, respectively. Interestingly, the $\alpha \gg 1$ near $P_c$ may enable the realization of the Fulde-Ferrel-Larkin-Ovshnikov (FFLO) states, or stabilize non-uniform superconducting states with spatial modulation order parameters and spin polarization [23, 27, 45].

**Conclusion**

In summary, we discovery the first 2D vdW heavy-fermion superconductor CeSiI near its AF QCP via the application of high pressure. The superconducting transition temperature exhibits a dome-shaped pressure dependence, with a maximum value of $T_c^{max} \approx 240$ mK around 7 GPa, where the AF magnetic order just vanishes and the coherence temperature reaches its minimum value of $T^* \approx 20$ K. The close proximity of the superconducting state to the AF instability, the unusually large $B_{c2}(0)$ exceeding 4-7 times the Pauli paramagnetic limit, and the enhancement of the electronic effective mass near the AF QCP all consistently indicate an unconventional superconducting mechanism. Our discovery opened a new avenue for studying the emergent novel quantum states in the 2D vdW heterostructure heavy-fermion systems.

**Methods**

**Crystal growth and characterizations** CeSiI single crystals were grown by a high-temperature-solid-state reaction as described in Refs. 1-2. The handling of the elements and compounds was all performed in an argon-filled glovebox. $CeI_3$ was first prepared by reaction of stoichiometric amounts of Ce and $I_2$ at 1000°C for three days. Then, the Ce pieces (99.95% purity from Zhong Nuo Advanced Material (Beijing) Technology Co., Ltd), Si powder (99.95% purity from Zhong Nuo Advanced Material (Beijing) Technology Co., Ltd) and the synthesized $CeI_3$ powder with a molar ratio of 2:3:1 were put into the Ta tube, which was sealed by using the arc melting method in an argon atmosphere. The Ta tube was subsequently sealed into a quartz tube under vacuum, before finally being placed into a box furnace. The furnace temperature was raised slowly to 1273 K for one day and then held for another six days. After that, the furnace was slowly cooled down to 973 K at the rate of 3 K/h and then turned off. Shiny, black plate-like hexagonal crystals with the lateral dimensions up to 2 mm can be obtained from the growth. The CeSiI single crystals are sensitive to air and moisture and thus are stored in an Ar-filled glovebox to avoid degradation. Detailed characterizations of CeSiI single crystals are shown in Extended Data Fig. 1. The energy-dispersive X-ray spectroscopy (EDS), which confirms that it contains three elements Ce, Si, I with the average molar ratio close to 0.97(5):1.03(1):1, which is close the chemical formula of CeSiI.

**High-pressure transport measurements** Temperature-dependent resistivity $\rho(T)$ under various pressures up to 14 GPa was measured by using a standard four-probe configuration in a palm-type cubic anvil cell (CAC) apparatus [49, 50], which can generate excellent hydrostatic pressures due to three-axis compression geometry and the adoption of liquid pressure transmitting medium (PTM). The electrical current was applied within the *ab* plane and the external magnetic field was applied parallel to the *c* axis. The pressure values in CAC were estimated from the calibration curve predetermined via monitoring structural transitions of bismuth (Bi) and tin (Sn) metals at room temperature [50]. The low-temperature measurements were carried out in a $^4$He exchange-gas cryostat (down to 1.5 K) or a dilution refrigerator (DR) cryostat (down to 0.05 K) equipped with a 12 T superconducting magnet at the Synergic Extreme Condition User Facility (SECUF).

**Acknowledgements**

This work was supported by the National Key Research and Development Program of China (Grants No. 2023YFA1406100, 2024YFA1408400, 2023YFA1607400, 2023YFA1406500, 2022YFA1403800, 2022YFA1402203), the National Natural Science Foundation of China (Grant No. 12025408, 12474055, 12174424, 12522407, 12404067, U23A6003, 12274459, 12174429), the Strategic Priority Research Program of CAS (Grant No. XDB33000000), the CAS President's International Fellowship Initiative (Grant No. 2024PG0003), the Outstanding Member of Youth Promotion Association of Chinese Academy of Sciences (Grant No. Y2022004) and the CAS Superconducting Research Project (Grant No. SCZX-0101). Part of this work was supported by JSPS KAKENHI 25K00951. The high-pressure measurements were performed at the CAC station of Synergetic Extreme Condition User Facility (SECUF, https://cstr.cn/31123.02.SECUF).

## Author contributions

J.C., B.W. and H.L. supervised this project. W.L. and H.L. synthesized the CeSiI single crystals, characterized their structure using XRD and EDX and then measured magnetic properties at ambient pressure; T.S., Q.D., B.W., J.C. performed resistivity and ac magnetic susceptibility measurements at ambient pressure and high pressures by using the cubic anvil cell apparatus with the supports of P.Y., H.M., Z.T., N.W., J.S. and Y.U.; Y.Y. gave important advice from a theoretical perspective; J.C., B.W., Y.Y, and H.L. wrote the paper with inputs from all co-authors.

**Competing interests**

The authors declare no competing interests.

**Online content:**

Any methods, additional references, Nature Portfolio reporting summaries, source data, extended data, supplementary information, acknowledgements, peer review information; details of author contributions and competing interests; and statements of data and code availability are available at XXX.

## Figure Captions

**FIG. 1. Crystal structure and high-pressure resistivity of CeSiI.** (a, b) Crystal structure from side and top views, respectively. The $\rho(T)$ of CeSiI (S1) under pressures: (c) 2-6 GPa; (d) 6-11 GPa. The black and blue arrows mark the characteristic temperatures $T^*$ and $T_N$ for the formation of the Kondo coherence state and the AF order, respectively.

**FIG. 2. Superconducting properties of CeSiI.** (a) Low-$T$ $\rho(T)$ of CeSiI (S1) under various pressures from 6 to 11 GPa and temperatures down to 0.05 K. (b) Low-$T$ $\rho(T)$ of CeSiI (S1) under different magnetic fields at fixed pressures of 7, 7.5 and 8 GPa; the dashed line represents the linear extension of normal-state resistivity. (c) The $\rho(B)$ at 0.08 K for 7 and 7.5 GPa. (d) The $B_{c2}(T)$ and corresponding fitting curves using the WHH model marked by the dashed lines. In (a)-(b), the blue and red arrows mark the onset and zero-resistivity superconducting temperatures, $T_c^{onset}$ and $T_c^{zero}$, which are defined as the temperatures where $\rho(T)$ starts to deviate from the extrapolated normal-state behavior and reaches zero, respectively. In (d), we choose $T_c^{onset}$ to mark the superconducting transition and plot the $B_{c2}(T)$.

**FIG. 3. Temperature-pressure diagram of CeSiI.** Pressure dependence of the Kondo coherence temperature $T^*$, AF transition temperature $T_N$ and superconducting transition temperatures $T_c$. The characteristic temperatures of CeSiI from Ref.2 are included for comparison. The color contour plot of the exponent $n(T, P)$ was superimposed on the *T-P* diagram highlighting the crossovers from Fermi-liquid to non-Fermi-liquid behaviors. The white dashed points indicate the Fermi liquid boundary.

**FIG. 4. Evidence of quantum criticality.** (a) The $\rho$ vs $T^2$ plot of low-$T$ $\rho(T)$ data for CeSiI (S1). Pressure dependencies of (b) the residual resistivity $\rho_0$ and (c) the quadratic-temperature coefficient $A$ (left) and the $B_{c2}/B_P$ (right); $A(P)$ in (c) was extracted from the linear fitting to $\rho$ vs $T^2$ in (a) and fitted via $A(P) \propto (P-P_c)^{-\beta}$ shown as the solid lines. (d) Pressure dependencies of the exponent $n(P)$. All the dashed lines in (b) and (d) are guides for the eyes.

**Figure 1**

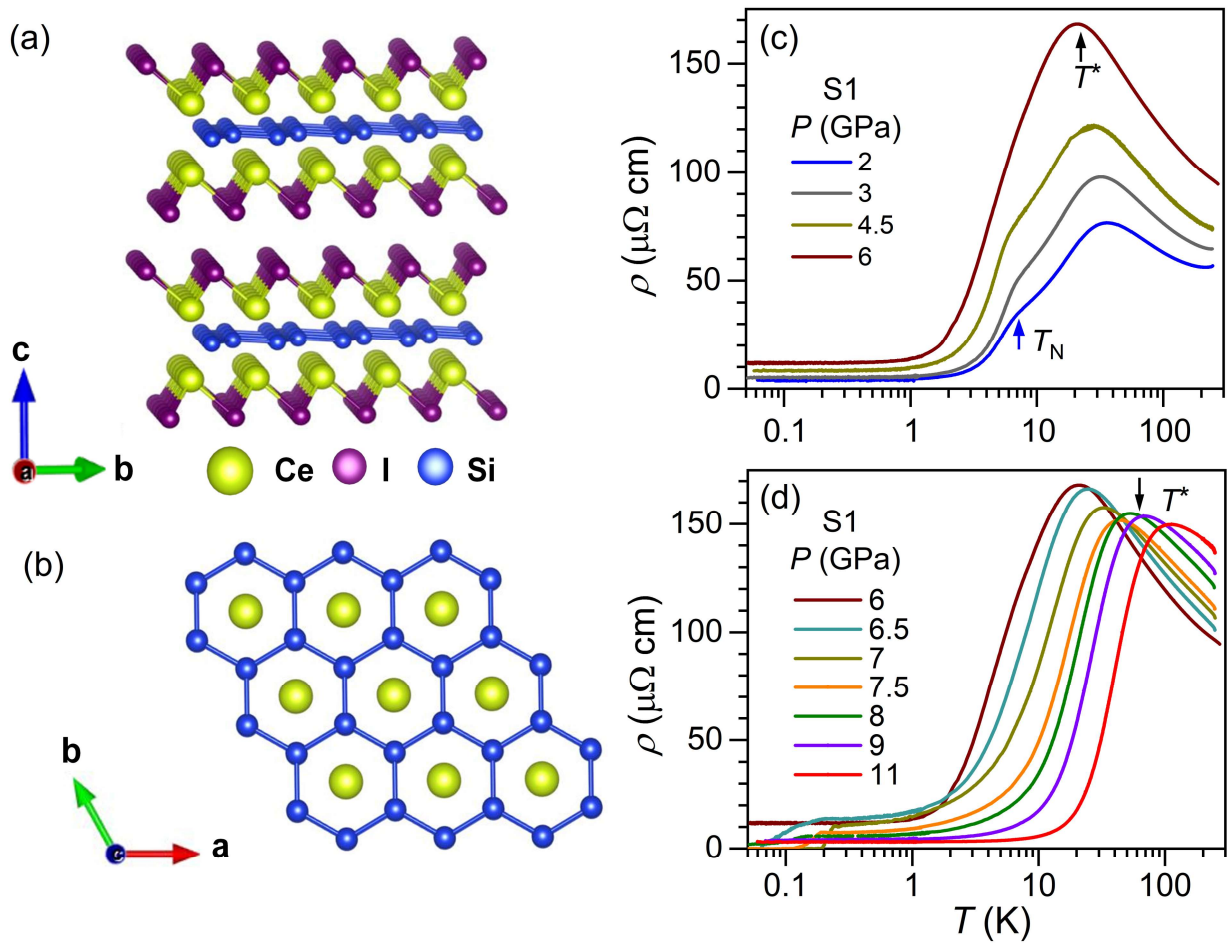

**Figure 2**

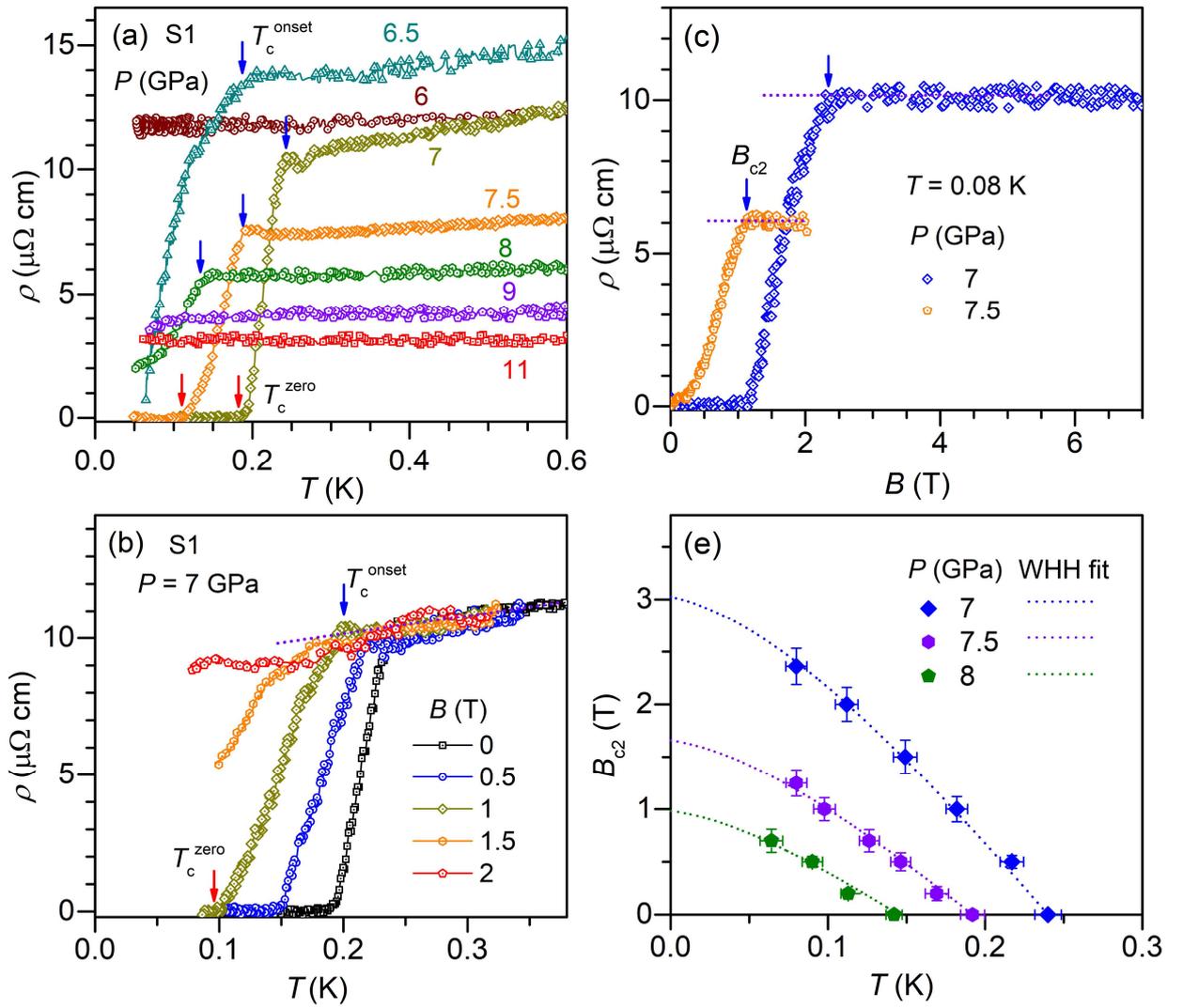

**Figure 3**

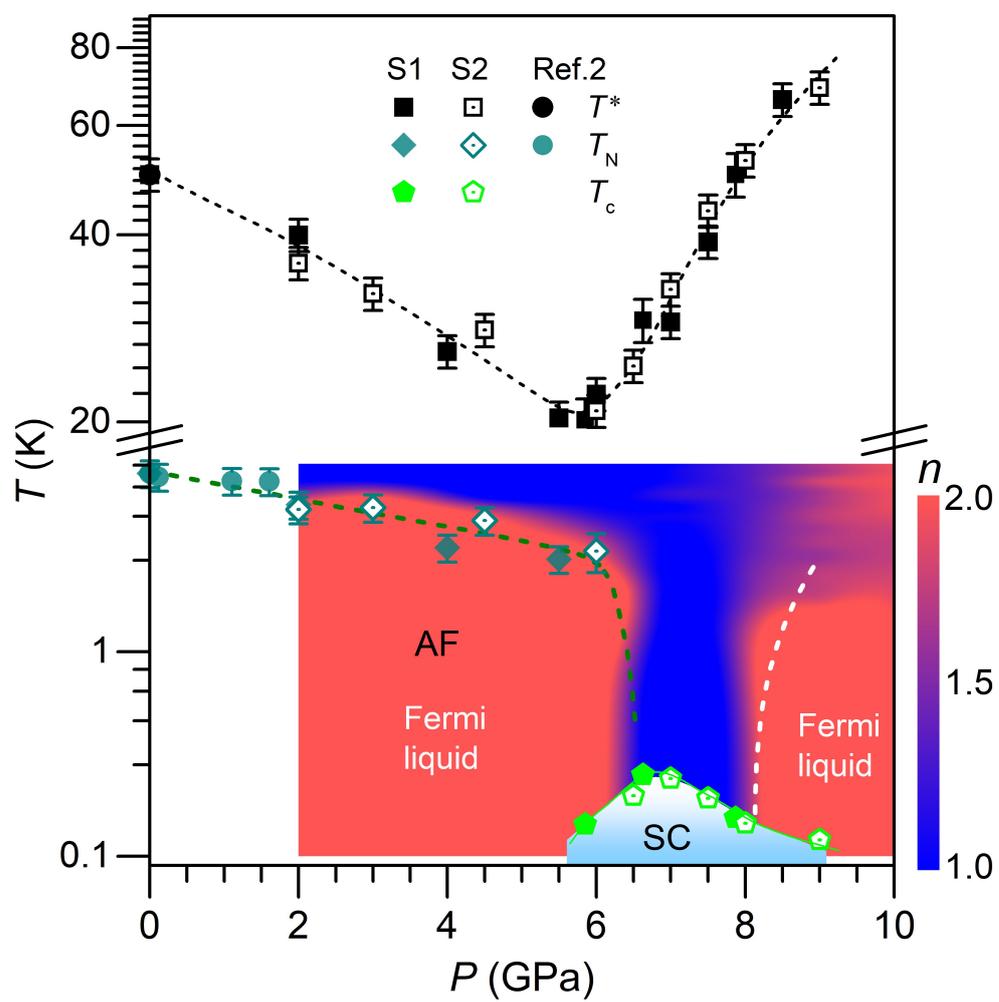

Figure 4

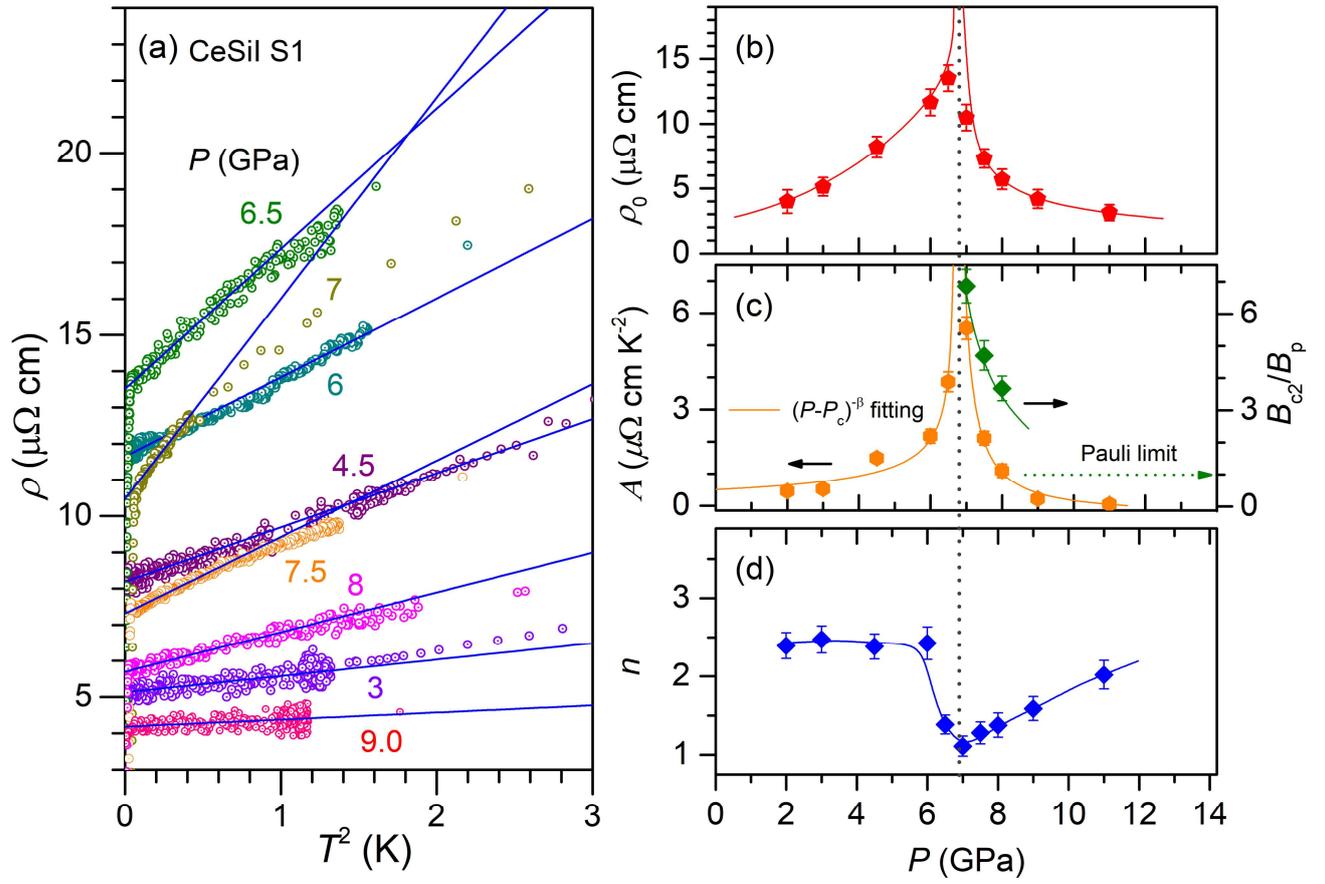